\documentclass[prl,a4paper,twocolumn,amssymb]{revtex4}
\usepackage{graphicx}

\begin {document} 

\title{Vortex shedding from a microsphere oscillating in superfluid $^4$He at mK temperatures and from a laser beam moving in a Bose-Einstein condensate}
\thanks{Dedicated to Klaus Dransfeld on the occasion of his 95th birthday}

\author{W. Schoepe}

\affiliation{Fakult\"at f\"ur Physik, Universit\"at Regensburg, D-93040 Regensburg, Germany}




\begin{abstract}
Turbulent drag of an oscillating microsphere, that is levitating in superfluid $^4$He at mK temperatures, is unstable slightly above a critical velocity amplitude $v_c$. The lifetime $\tau$ of the turbulent state is determined by the number $n$ of vortices shed per half-period. It is found that this number is identical to the superfluid Reynolds number. The possibility of moving a levitating sphere through superfluid $^3$He at microkelvin temperatures is considered. A laser beam moving through a Bose-Einstein condensate (BEC) (as observed by other authors) also produces vortices in the BEC. In particular, in either case a linear dependence of the shedding frequency $f_v$ on $\Delta v = v - v_c$ is observed, where $v$ is the velocity amplitude of the sphere or the constant velocity of the laser beam above $v_c$ for the onset of turbulent flow: $f_v = a \,\Delta v$, where the coefficient $a$ is proportional to the oscillation frequency $ \omega $ above some characteristic frequency $\omega_k$ and assumes a finite value for steady motion $\omega \rightarrow 0$. A relation between the superfluid Reynolds number and the superfluid Strouhal number is presented that is  different from classical turbulence.
\end{abstract}

\maketitle 


\section*{Introduction}
Quantum turbulence is a common phenomenon in superfluids, ranging from the dense $^4$He and $^3$He liquids to the very dilute Bose-Einstein condensates (BEC). The vortices have a quantized circulation $ \kappa $ = $h/m$, where $h$ is Planck's constant and $m$ is the mass $m_4$ of a $^4$He atom, or $2m_3$ of a Cooper pair in superfluid $^3$He, or an atom of a BEC gas. In $^4$He we have $\kappa \approx$ 10$^{-7}$ m$^2$/s. Vortices can be created, e.g., by stirring the superfluids with a moving object or by rotation. In the helium superfluids the easiest way to produce vorticity is by using oscillating objects like spheres, tuning forks or vibrating wires. Because of the simple geometry of a sphere its behavior is more transparent and more easily analyzed than that of the more complicated oscillating structures.

\noindent In the case of a BEC the moving object is typically a laser beam that presents an obstacle to the condensate. The laser beam is swept continuously through the condensate. In addition to the experiments there is a large number of theoretical work on the transition to turbulence based on numerical solutions (mostly 2-dimensional) of the nonlinear Schr\"odinger equation, often known as Gross-Pitaevskii equation, which is applicable for BECs but not for the dense helium liquids.

\noindent The motivation for the present article is partly a comparison of the frequencies at which the vortices are shed in both types of superfluids. We find that in spite of the very different experimental parameters (density, coherence length, speed of sound, interaction strength, linear dimensions, etc.) the shedding frequencies are similar. Moreover, the change from oscillatory flow to steady flow will be discussed for both superfluids. This article is an update of an earlier review \cite{review,Add}, and it is an outlook to more experiments in the future. It is written for a more general readership than the one for the earlier review.

\section*{The resonator}
The experimental technique makes use of superconducting levitation of a ferromagnetic sphere (radius $R$ = 0.12 mm, mass $m$ = 27$\mu$g) between superconducting niobium electrodes of a horizontal parallel plate capacitor (spacing $d$ = 1 mm), see Fig.1 (for details, see\cite{review}).

\begin{figure}[h] 
\centerline{\includegraphics[width=0.7\linewidth]{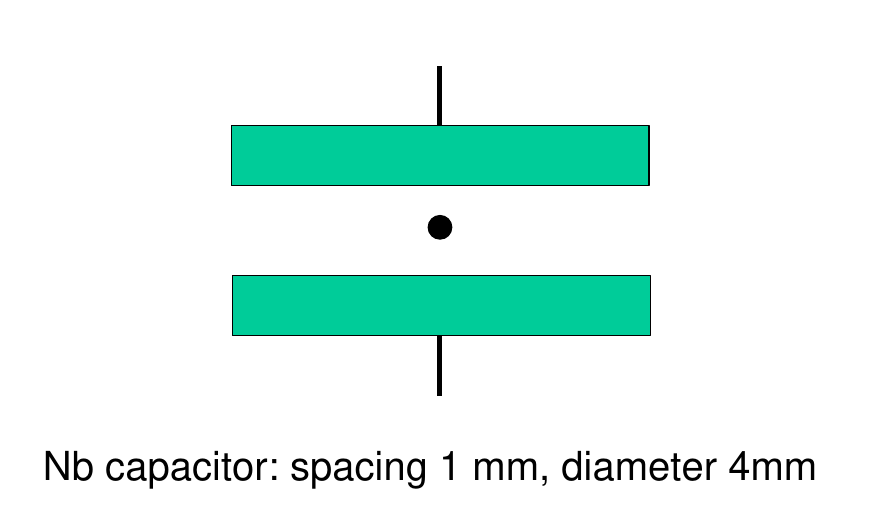}}
\caption{Schematic of our resonator based on superconducting levitation of a magnetic sphere carrying an electric charge in a niobium capacitor filled with superfluid $^4$He at mK temperatures (Color figure online)}
\label{fig:1}       
\end{figure}
\noindent We also have tested capacitors made of the high-$T_c$ superconductor YBCO, both bulk and thin film \cite{YBCO}. At low oscillation amplitudes the quality factors in vacuum were as high as for the one made of niobium ($\sim$ 10$^{6}$), but at larger amplitudes the damping in vacuum became nonlinear, in contrast to the results with the Nb capacitor. We therefore preferred to use the latter one.

\noindent Before cooling the capacitor into the superconducting state (for Nb at 9.2 K), we apply several hundred volts to the bottom electrode charging the sphere to about $q \sim $ 1 pC. 
Vertical oscillations around the equilibrium position of the levitating sphere can be excited by applying an ac voltage $U_{ac}$ at resonance ($\sim$ 120 Hz) in the range from 0.1 mV to several volts, exerting on the sphere a driving force $F = q\,U_{ac}/d$. The oscillations induce an ac current $I = q\,v/d$ that is detected by an electrometer. By measuring $v(F)$ and the exponential free decay of the oscillations at the same temperature we determine the electric charge. The stability of $q$ is checked every morning by repeating some of the data from the day before. Usually the charge is found to be quite stable.

\noindent Because no mechanical support is needed, the sphere moves at a well-defined velocity. Moreover, the simple spherical geometry makes the data transparent and more directly accessible in a quantitative way, in particular the laminar and the turbulent drag forces on the sphere can be identified quantitatively, as will be shown below.

\section*{Experimental results}
The velocity amplitude as a function of the driving force at 300 mK is displayed in Fig.2.

\begin{figure}[h]
\centerline{\includegraphics[width=1.1\linewidth]{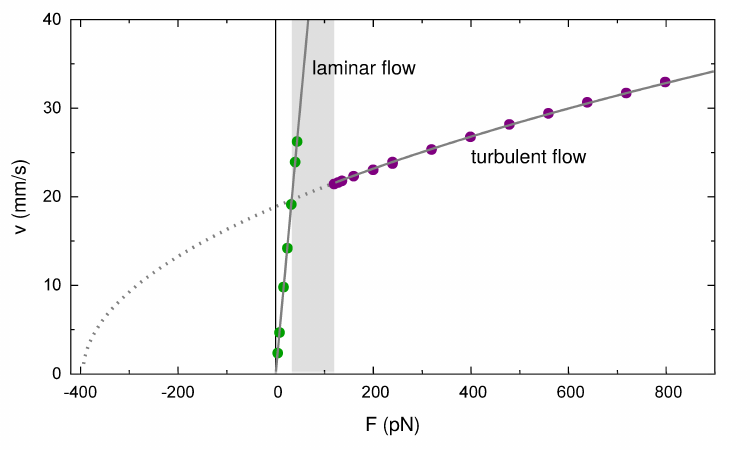}}
\caption{(From \cite{NiemetzSchoepe2004}) Velocity amplitude as a function of the driving force amplitude at 300 mK, oscillation frequency 114 Hz. There are three different regimes: At small drives the linear increase is the regime of potential flow and the slope is given by ballistic phonon scattering; at larger driving forces we observe stable nonlinear turbulent drag; and the shaded area indicates an unstable regime slightly above a critical velocity where the flow switches intermittently between both patterns, see Fig.3 below. (Color figure online)}
\label{fig:2}
\end{figure}

\noindent In the linear regime the drag force is a linear function of the velocity, namely $F = \lambda (T) \, v$, and the coefficient $\lambda $ is attributed to ballistic phonon scattering:
\begin{equation}
\lambda(T)\,=\,\rho_{\mathsf{ph}}  \cdot c \cdot \pi R^2\,\propto T^4\,,\label{Eq:1}
\end{equation}
where the phonon density $\rho_{\mathsf{ph}}$ rapidly varies as $T^4$, and $c$ is the velocity of sound. The quantitative agreement of Eq.(1) with the linear data in Fig.2 is a testbed, reassuring us that our technique yields understandable and reproducible results.  

\noindent The nonlinear dependence of $v(F)$ in Fig.\,\ref{fig:2} can be properly described by a quadratic drag force $F_D \propto (v^2 - v_c^2) $. In contrast to a classical liquid, the apex of the parabolic shape of $v(F)$ is shifted to the left of the origin by 0.4 nN, and the resulting finite intercept at $F$ = 0 indicates a velocity range of frictionless flow which is the paradigm of superfluidity. 
Moreover, we observe a sharp onset of the turbulent regime, whereas in a classical liquid there are about three orders of magnitude in flow velocity between Stokes' regime of laminar flow and fully developed turbulence, where the classical turbulent drag on a sphere is given by $\gamma v^2$ with $\gamma = c_D \rho \pi R^2 / 2 $, where $\rho $ is the density of the liquid and the drag coefficient $c_D$ of a sphere is approximately 0.4 \cite{LL}. 

\noindent Of particular interest is the unstable regime, where in a narrow interval from the critical velocity $v_c$ up to $v_c + \Delta v $, where $\Delta v/v_c \le$ 0.03, the flow switches intermittently between turbulence and potential flow, see Fig.3.

\begin{figure}[h]
\centerline{\includegraphics[width=1.1\linewidth]{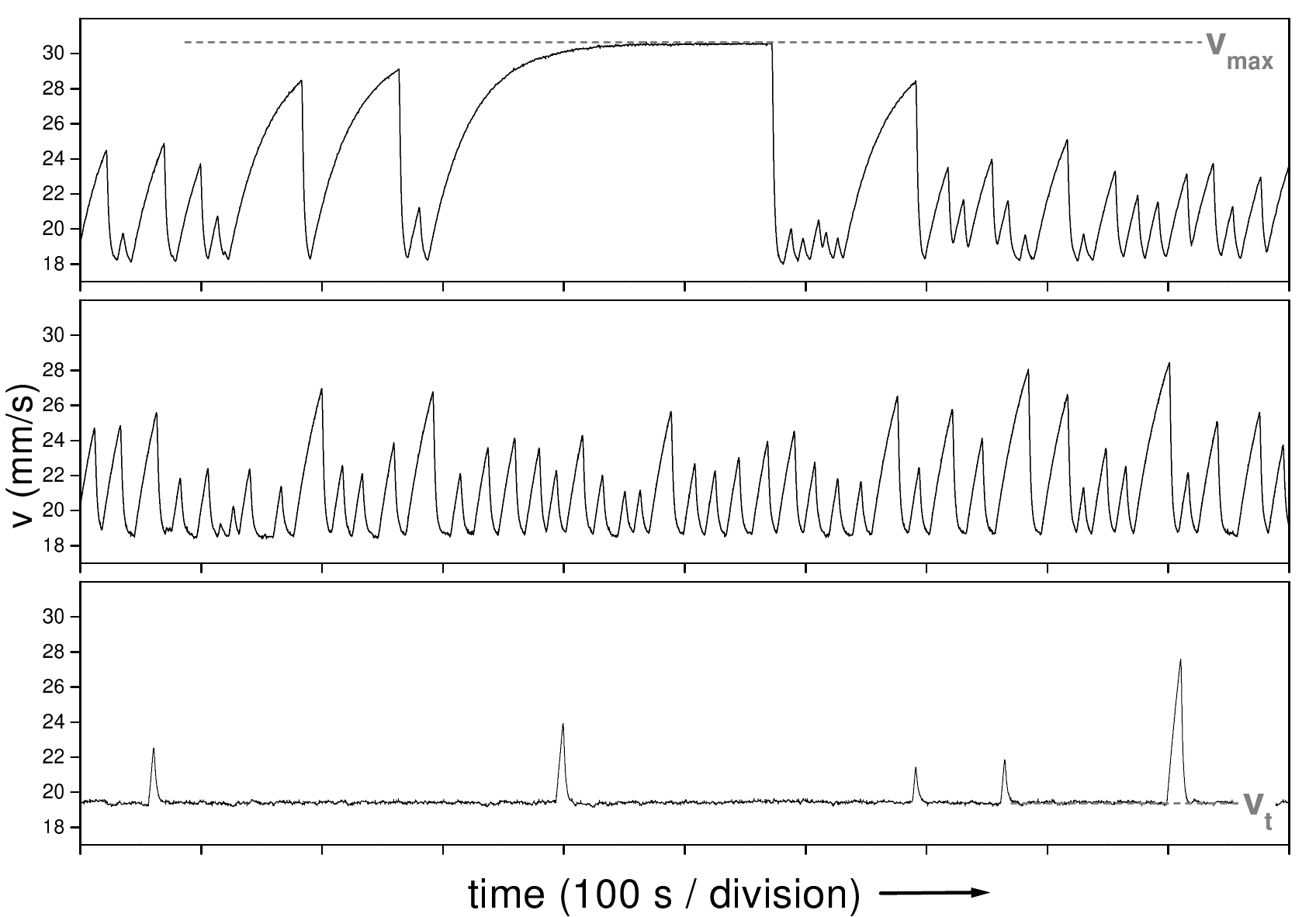}}
\caption{(From \cite{JLTP2002}) Three time series of the velocity amplitude at 300 mK and at three different driving forces (in pN: 47, 55, and 75), are shown from top to bottom. The low level $v_t$ corresponds to turbulent flow while the increase occurs during a laminar phase. With increasing drive the lifetimes of the laminar phases become shorter, whereas the lifetimes of turbulent phases grow rapidly. The time interval shown here extends over 1000 s $\approx $ 17 min, oscillation frequency 114 Hz}
\label{fig:3}
\end{figure}

\noindent We find that the lifetimes $t$ of the turbulent phases are exponentially distributed exp$(-t/\tau )$, and the mean lifetimes $\tau $ increase rapidly with the driving force, namely as 
\begin{equation}
\tau (F) = \tau_0\,\exp[\,(F/F_1)^2], \label{Eq:2} 
\end{equation}
see Fig.4.
\begin{figure}[h]
\centerline{\includegraphics[width=1.4\linewidth]{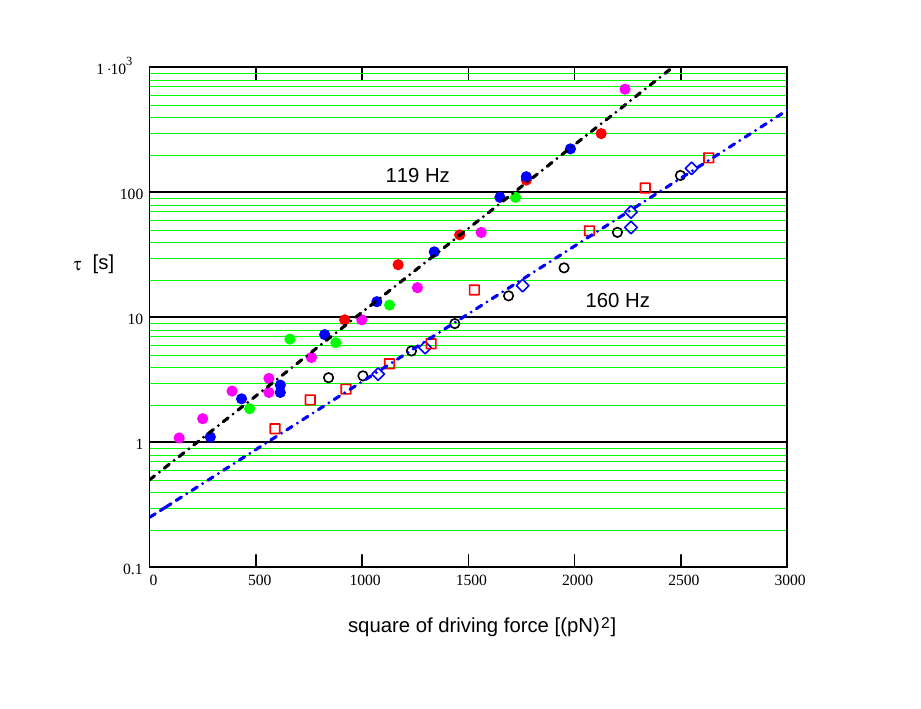}}
\caption{(From \cite{JLTP2013}) Mean turbulent lifetimes as a function of the driving force at two different oscillator frequencies. Each data point is obtained from a time series, some of which lasted up to 36 hours.
The straight lines are fits of Eq.(2) to the data. The data of the 119 Hz oscillator were taken at 4 different temperatures (in mK): red 403; blue 301; green 200; and violet 100. The data at 160 Hz were taken at 300 mK (black circles); at 30 mK with a mixture of 0.05\% $^3$He (red squares); at 30 mK with 0.5\% of $^3$He (blue diamonds). Note that the slopes $1/F_1^2$ and the intercepts $\tau_0$ are independent of temperature and $^3$He concentration, but both depend on the oscillation frequency. (Color figure online)}
\label{fig:4}       
\end{figure}
\noindent The fitting parameters are $\tau_0$ = 0.5 s at 119 Hz and 0.25 s at 160 Hz, and $F_1$ = 18 pN and 20 pN, respectively. The force $F_1$ can be interpreted as being caused by the loss of kinetic energy of the sphere due to the shedding of {\it one} vortex ring of radius $R$ during {\it one} half-period \cite{JLTP2013}. From dimensional arguments and a fit to the data we find
\begin{equation}
F_1 = 1.3 \rho \kappa R \sqrt{\kappa \omega}, 
\end{equation}
\noindent where $\rho$ is the density of the liquid and $\omega = 2\pi f$. The drag force is obtained from the data $v(F)$ and is given by 
\begin{equation}
F_D(v) = (8/3\pi)\gamma (v^2 - v_c^2).
\end{equation}
\noindent The numerical factor $8/3\pi$ = 0.85 takes into account the energy balance for an equilibrium oscillation amplitude: energy gain from the drive and loss from a quadratic damping must cancel. While Eq.(4) is deduced from the experiment up to velocities of ca.\,100 mm/s, which is 5 times larger than $v_c$, Eq.(3) is proven valid only in the small interval $\Delta v/v_c \le$ 0.03 where $\tau $ was measurable. In this regime we may approximate Eq.(4) by  
\begin{equation}
F_D(v) = (8/3\pi)\,2\gamma \, v_c\,\Delta v.
\end{equation}

\noindent We assume that the number $n = F_D/F_1$ is the average number of vortex rings emitted per half-period. Inserting Eq.(3) and Eq.(5), and using our results $v_c = 2.8 \sqrt{\kappa \omega}$ [1], we find
\begin{equation}
n = \frac{F_D}{F_1} = \frac{(8/3\pi)\,2 \gamma  v_c \, \Delta v}{1.3\,\,\rho \,\kappa \,R\, \sqrt{\kappa \,\omega}} = \frac{\Delta v}{v_1}, 
\end{equation}
where $v_1 = 0.48\,\kappa /R$ = 0.39 mm/s, and $n$ lies in the interval 0.7$<n<$3.0.\\ 
\noindent In Fig.5 we plot the normalized mean lifetime  
\begin{equation}
\tau^*(\Delta v) \equiv \tau /\tau_0 = \exp{[(\Delta v/v_1)^2]}. 
\end{equation}
The salient feature is that $\tau^*$ is independent of the oscillation frequency, of the temperature, and is not affected by $^3$He impurities. It is remarkable that only $\Delta v $ matters and not the frequency $\omega $ directly. The only explicit frequency dependence is in $\tau_0$. A theoretical interpretation of $\tau^* ( \Delta v)$ is, to our knowledge, presently not available. We note that the lifetimes of the turbulent phases increase faster than exponentially with $\Delta v$ or with the superfluid Reynolds number. This characteristic is called "supertransient chaos", here for the first time in a superfluid \cite{Tel}.

\begin{figure}[h]
\centerline{\includegraphics[width=1.2\linewidth]{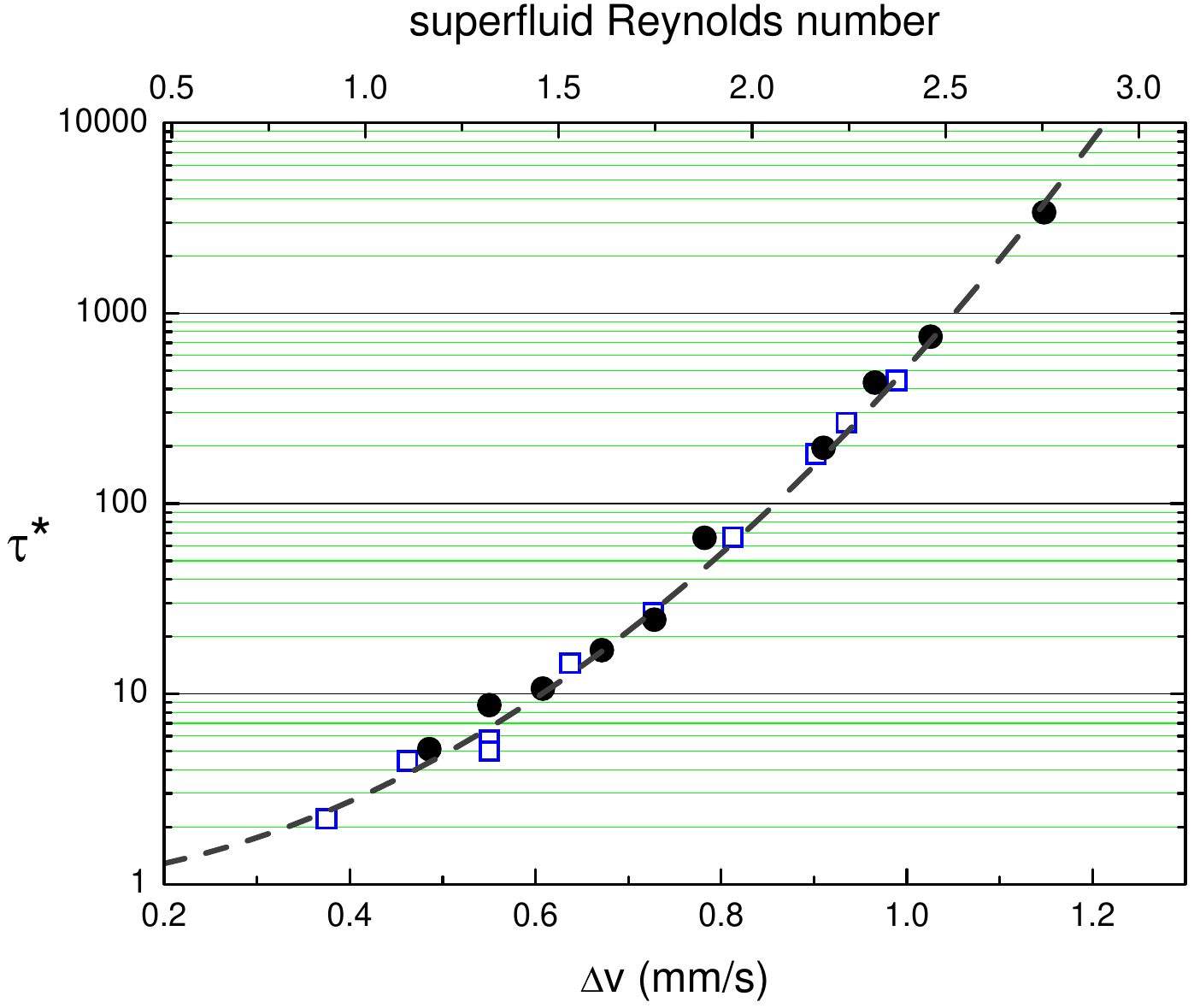}}
\caption{(From \cite{JETP}) The normalized lifetimes $\tau^* = \tau/\tau_0$ as a function of $\Delta v = v - v_c$ for the 119 Hz oscillator at 301 mK (blue squares) and the 160 Hz oscillator at 30 mK  with 0.05\% $^3$He (black dots). Note the rapid increase of $\tau ^*$ by 3 orders of magnitude over the small velocity interval of ca. 0.7 mm/s. At the top axis, the corresponding values of the superfluid Reynolds number are given, see text below. The dashed line is calculated from Eq.(7) (Color figure online)}
\label{fig:5}       
\end{figure}

\section*{The superfluid Reynolds number}

In classical hydrodynamics the Reynolds number is defined as $ Re = v\,D/\nu $ where $D$ is a characteristic length scale, and $\nu $ is the kinematic viscosity. Since there is no viscosity in superfluid helium at mK temperatures, $\nu $ is replaced by the circulation quantum $\kappa $ that has the same dimension as $\nu $. Hence, $Re_s = v\,D/\kappa $ describes a circulation in units of the quantum $\kappa $ \cite{JETP}. However, because below the critical velocity there is no turbulence, it is clearly appropriate to modify the superfluid Reynolds number as 
introduced by Reeves et al. \cite{Reeves}, namely 
\begin{equation}
  Re_s = (v - v_c)\,D /\kappa.
\end{equation}
Applying this definition to our case \cite{arxiv}, we choose $D = 2 R$ as the characteristic length scale and define $v_0 \equiv \kappa /2\,R$, we have ,
\begin{equation}
Re_s = \frac{\Delta v}{v_0},
\end{equation}
where in our case $v_0 = 0.40$ mm/s. 
This result is valid for a sphere, but no assumptions have been made concerning the dimension of the flow (2D or 3D)  nor of its type (steady or oscillatory). We note, that the ratio $\kappa /R$ determines the self-induced velocity of a vortex ring of radius $R$.

\noindent Comparing Eq.(6) with Eq.(9) we note that $v_0$ and $v_1$ differ only by 4 \%. From the accuracy of the numerical factors of $v_c$ and $F_1$ we estimate an uncertainty of $n$ in Eq.(6) to be about 10\%, i.e., within our experimental resolution we have
\begin{equation}
Re_s = n.
\end{equation} 
This is a surprisingly simple result. \\

\noindent It should be mentioned that in simulations of vorticity in 2D, a similar result has been calculated, namely that the superfluid Reynolds number is given by the number of 2D vortices \cite{reeves2}.

\section*{Outlook: A levitating sphere moving in superfluid $^3$He?}
Although there is a large body of literature on oscillating structures in superfluid $^3$He, like vibrating wires, tuning forks or grids, but so far there are no experiments with a floating sphere, that could be compared in detail with our work on $^4$He . However, very recently a fascinating attempt has been suggested by the group of D. Zmeev at Lancaster \cite{Zmeev}. Due to the large magnetic fields that are required for adiabatic demagnetization in order to cool liquid $^3$He into the superfluid state, our design cannot be used. Instead the authors suggest to levitate a superconducting sphere (radius $R$ = 0.55 mm) by a set of coils. This design will offer both oscillations as well as steady motion of the sphere. Moreover, the sphere will consist of a hollow plastic body covered with a thin film of indium. Therefore, its surface can be expected to be much smoother than that of our ferromagnetic particle. It will be very interesting to compare their results with ours. Because the radius $R$ is now 4.4 times larger than that of our sphere, all quantities that depend on $R$ will be different, e.g., $F_1$ (Eq.(3)), $v_1$ (Eq.(6)), $\tau $ (Eq.(2)). In addition all quantities depending on the oscillation frequency $\omega $, e.g., the critical velocity $v_c$ \cite{review}, can  easily be varied, in contrast to our work, where $R$ was fixed and $\omega $ could only be changed by a new levitation status (due to flux frozen in the electrodes), when warming the measuring cell above $T_c$ of Nb. And finally, the physics of superfluid $^4$He is very different from that of superfluid $^3$He, where exotic surface states exist, which very likely may affect the motion of the sphere. The results will definitely be fascinating.
\section*{Vortex shedding from the sphere and from a laser beam moving through a BEC}

In this Section we compare
our own experiments in superfluid helium as described above with those in a BEC as observed by other authors, where a moving laser beam sheds vortices above a critical velocity. In particular, the frequency $f_v$, with which vortices are shed, is found to be similar.\\
 
\noindent Beginning with our experiments, we obtain from the average number $n$ of vortex rings shed per half-period, see Eq.(6), the shedding frequency
\begin{equation}
f_v =  2 n f = \frac{2f \Delta{v}}{v_1} = a\,\,\Delta v, 
\end{equation}
\noindent where the coefficient $a$ = 2$f /v_1$. At $f$ = 119 Hz we obtain $a$ = 0.60 $\mu$m$^{-1}$ and at 160 Hz $a$ = 0.80 $\mu$m$^{-1}$, and shedding frequencies $f_v$ ranging up to $\sim $ 500 s$^{-1}$. The linear increase of $a(f)$ must change to some finite limit when $f \to 0 $, because it is clear that vortices can be shed also for steady motion. 

\noindent From $1/a$ we have a characteristic length scale, which is given here by $v_1 / 2f$. At $f$ = 119 Hz we obtain $1/a$ = 1.7 $\mu$m and at 160 Hz $1/a$ = 1.3 $\mu$m. This length can be interpreted as the distance a vortex ring travels during one half-period. In the case of steady motion we postulate that the characteristic length scale is given by the radius $R$ of the sphere. In our case this is a small but finite value $a \sim 1/R $ = 0.008 $\mu$m$^{-1}$. In Fig.6 a schematic of the frequency dependence of $a$ is shown. 

\begin{figure}[ht]
\centerline{\includegraphics[width=1.1\linewidth]{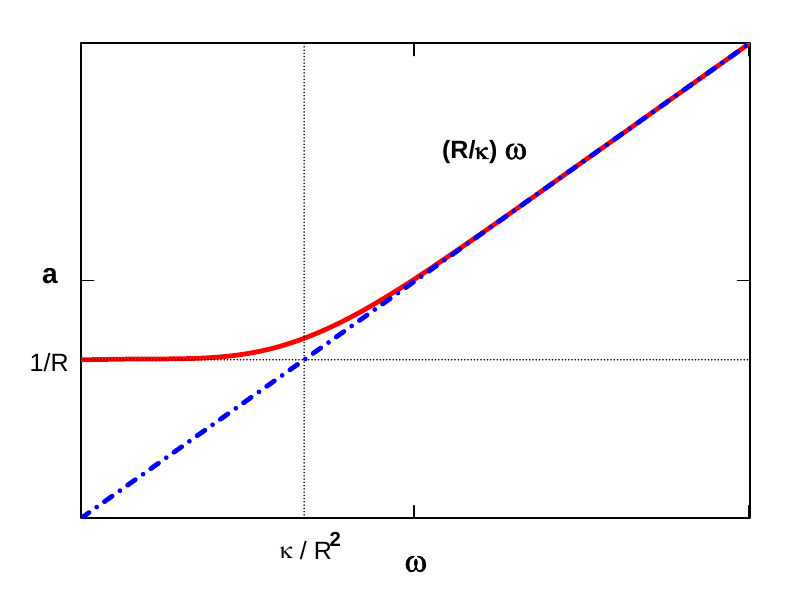}}
\caption{(From \cite{JLTP2017}) Sketch of the coefficient $a$ of $f_v = a\, \Delta v $ as a function of the oscillation frequency $\omega = 2\pi f.$ At small frequencies the radius $R$ of the sphere is taken as the characteristic length scale, hence $a \sim 1/R$, whereas at large frequencies $a$ scales as $(R/\kappa)\,\omega$, see Eq.(11). The characteristic frequency that marks the transition between both regimes is given by $\kappa /R^2$. Numerical factors of order 1 are neglected. (Color figure online)}
\label{fig:6}       
\end{figure}

\noindent 
Shedding of vortex dipoles in a stable and periodic manner has been observed recently at Seoul National University by moving a repulsive Gaussian laser beam steadily through a BEC of $^{23}$Na atoms \cite{Shin}. The shedding frequency $f_v$ is also given by $a\, \Delta v $, where now $v_c$ = 0.99 mm/s and $a$ = 0.25 $\mu$m$^{-1}$. Because the beam was moved steadily, we assume that the relevant length scale is given by the radius of the beam $R$ = 4.6 $\mu$m, in
accordance with the arguments presented above. Therefore, we estimate $a \sim 1/R $ = 0.22 $\mu$m$^{-1}$, in fair
agreement with the experimental result, and $ v_c \sim \kappa /R $ = 3.7 mm/s (where $\kappa $ = 1.7 $10^{-8}$ m$^2$/s for the $^{23}$Na BEC). If the beam would have been oscillating at a frequency substantially 
larger than $\omega_k$ = $(\pi /2)\,\kappa/R^2$ = 1.26 $10^3$ s$^{-1}$ (or 201 Hz), we would expect a shedding frequency $f_v$ proportional to $ \omega $, in accordance with our results in $^4$He. Apparently this has not yet been investigated.\\
\section*{The superfluid Strouhal number}
 
In classical time dependent flows the Strouhal number $Sr$ is important in addition to the Reynolds number. Flows are said to be similar when both of their numbers are the same. There is a weak dependence of the Strouhal number on the Reynolds number \cite{Roshko}. $Sr$ is related to the vortex shedding frequency $f_v$, namely \cite{st}
\begin{equation}
Sr \equiv f_v \,(2R/v), 
\end{equation}
\noindent
where in our case $f_v$ = $a\, \Delta v $,
see Eq.(11).\\ 

\noindent
At low frequencies $0 \le\omega < \omega_k $ we have $ a $ = $1/R $, hence $f_v $ = $ \Delta v /R $  and $Sr$ = $2\,  \Delta v/v$, i.e., $Sr$ varies from 0 at $v_c$ asymptotically up to 2 for large velocities $\Delta v/v \rightarrow$ 1. However, because the available data on vortex shedding in a BEC range from $v_c$ up to $\approx$ 2$v_c$ \cite{Shin}, only the initial rise of $Sr$ can be inferred, where $Sr \le $ 1.
Moreover, using the superfluid Reynolds number in order to replace $\Delta v$ by $Re_s$ we find the relation
\begin{equation}
Sr = Re_s (\kappa /R v), \;(v\ge v_c).
\end{equation} 
\noindent
\noindent
For frequencies larger than $\omega_k $ we have from Eq.(11) $a$ = $2 R \omega /\pi \kappa$ and $Sr $ = $(4\, R^2 \omega/\pi \kappa)\Delta v/v $. In that case $Sr$ grows from 0 asymptotically to a maximum value $ Sr_{max} = 4\, R^2 \omega/\pi \kappa $, which for our values of $R$ and $\omega $ is $\sim $ 140. Using the characteristic frequency $\omega_k $ we may simply write $Sr_{max} = 2\, \omega /\omega_k$, where in our case $\omega_k $ = 10.9 s$^{-1}$ (or 1.74 Hz). Also in this case, only the initial rise from 
$v_c$, where $\Delta v/v_c \le$ 0.03, has been experimentally accessible, hence limiting $Sr \le$ 4. Finally, we find the relation
\begin{equation}
Sr = Re_s (2\,R\omega /\pi v), \; (v\ge v_c).
\end{equation}

\noindent
The above results are obtained from our data at 2 frequencies larger than $\omega_k $ and those  of Shin's group \cite{Shin} from vortex shedding by a laser beam moving steadily through a BEC. It would be interesting to have more data to compare with $a(\omega )$, see Fig.6. \\

Finally, we should like to mention that a recent article was published by Y. Lim et al. "Vortex shedding frequency of a moving obstacle in a Bose-Einstein condensate", containing results on the superfluid Strouhal number that are similar to ours \cite{Lim}.\\

\noindent
A modification of Eq.(6) for larger $ \Delta v/v_c $ in liquid helium will be considered in the next Section. \\ 

\section* {Vortex shedding from the oscillating sphere when $\mathbf {\Delta v/v_c \gg }$ 0.03}
In this regime, the linear approximation of the drag force Eq.(5), is not valid any more. Instead, it is necessary to modify Eq.(6) by inserting Eq.(4):

\begin{equation}
n = \frac{F_D}{F_1} = \frac{(8/3\pi) \gamma (v^2-v_c^2) }{1.3\,\rho \,\kappa \,R\, \sqrt{\kappa \omega}} = \frac{\Delta v}{v_1} \left(1+ \frac{\Delta v}{2v_c}\right).
\end{equation}

\noindent
Consequently, there is a quadratic term $(\Delta v)^2$ in the shedding frequency $f_v(\Delta v)$ = $2\,f\,n(\Delta v)$ and also in the superfluid Strouhal number $Sr$:

\begin{equation}
Sr = \frac{2f}{v_1}\left(1+\frac{\Delta v}{2v_c}\right)\, 2R\,\frac{\Delta v}{v}.
\end{equation}
Therefore, $Sr$ no longer saturates at large velocities, but instead it is proportional to $v$, and Eq.(14) changes to
\begin{equation}
Sr = Re_s (2\,R\omega /\pi v)(1+ Re_s v_1/v_c), \; (v\ge v_c).
\end{equation}
The term $Re_s v_1/v_c$ in Eq.(17) becomes dominant (i.e., reaches a value $\geq 1$) when $Re_s \geq v_c/v_1$. Using $v_1 = \kappa /2R$ we obtain $\Delta v \geq v_c$. Similarly, the second term in the brackets in Eq.(15) will be become dominant when $ \Delta v \geq 2 v_c$.

Because we do not have experimental data of $f_v$ for this regime, this is a purely theoretical consideration based on the assumption that the drag force $F_1$ due to the shedding of 1 vortex ring per half-period, see Eq.(3), remains valid.

\section*{Summary}
1. In the regime of potential flow our oscillating sphere shows the expected linear drag force due to ballistic phonon scattering with geometric cross section, because the wavelength of a thermal phonon in liquid helium at 0.1 K is about 0.1 $\mu$m, which is much smaller than the size of the sphere. The quantitative agreement of the data with ballistic phonon scattering is a testbed for our resonator.\\

\noindent 2. In the stable turbulent regime it is surprising that we find the classical turbulent drag $\gamma v^2$ of a sphere, except for a shift along the force axis $F(0)$ = $-(8/3\pi) \gamma v_c^2 \approx -$ 0.4 nN and $F$ = 0 for $v \leq v_c$. This result needs a theoretical explanation for turbulent oscillatory flow, in particular a rigorous calculation of $v_c$ as function of the oscillation frequency.\\

\noindent 3. In the regime of intermittent switching between laminar and turbulent flow, we find a lifetime of the turbulent states that can be attributed to the number $n$ of vortices that are shed per half-period with the same radius as the sphere. The lifetimes increase exponentially with $n^2$. We find that $n$ is identical to the superfluid Reynolds number. The frequency with which the vortex rings are shed increases linearly with $(v-v_c)$ both in our experiment and in a BEC (as observed by other authors), and the coefficient is determined by the radius of the sphere or of the laser beam, and the oscillation frequency.\\

\noindent 4. More experiments on vortex shedding in a BEC with a laser beam oscillating at various frequencies would be interesting to further test the frequency dependence $a (\omega )$ in Fig. 6.\\

\noindent
\textbf{Acknowledgements:}
Our data on $^4$He presented in this work were obtained together with my former students Hubert Kerscher and Michael Niemetz to whom I am very thankful for their wonderful co-operation. Helpful comments from A.L. Fetter (Stanford University) on superfluid vortex dynamics are gratefully acknowledged.

\bigskip
\noindent e-mail: wilfried.schoepe@ur.de\\




\end{document}